\documentclass[11pt,twoside]{article}
\usepackage{asp2014}
\aspSuppressVolSlug
\resetcounters
\usepackage[utf8]{inputenc}
\usepackage{import,graphicx,natbib}
\usepackage{amsmath,amssymb,upgreek}

\begin{document}
\title{Extreme Scattering Events and Symmetric Achromatic Variations}
\author{H.~K.~Vedantham
\affil{Cahill center for Astronomy and Astrophysics, California Institute of Technology, 1200 E. California Blvd., Pasadena, CA 91125, USA}}
\begin{abstract}
Radio variability in quasars has been seen on timescales ranging from days to years due to both intrinsic and propagation induced effects \citep{kell1968,rickett1990}. Although separating the two is not always straight-forward, observations of singular `events' in radio light curves have led to two compelling, and thus far unresolved  mysteries in propagation induced variability--- extreme scattering events (ESE) that are a result of plasma lensing by AU-scale ionized structures in the interstellar medium, and symmetric achromatic variability (SAV) that is likely caused by gravitational lensing by $\gtrsim 10^3\,M_\odot$ objects. Nearly all theoretical explanations describing these putative lenses have remarkable astrophysical implications. In this chapter we introduce these phenomena, state the unanswered questions and discuss avenues to answer them with a $\sim $weekly-cadence flux-monitoring survey of $10^3-10^4$ flat-spectrum radio quasars with the ngVLA.
\end{abstract}

\section{Extreme scattering events}
\begin{figure*}
\centering
\begin{tabular}{ll}
\includegraphics[width=0.47\linewidth]{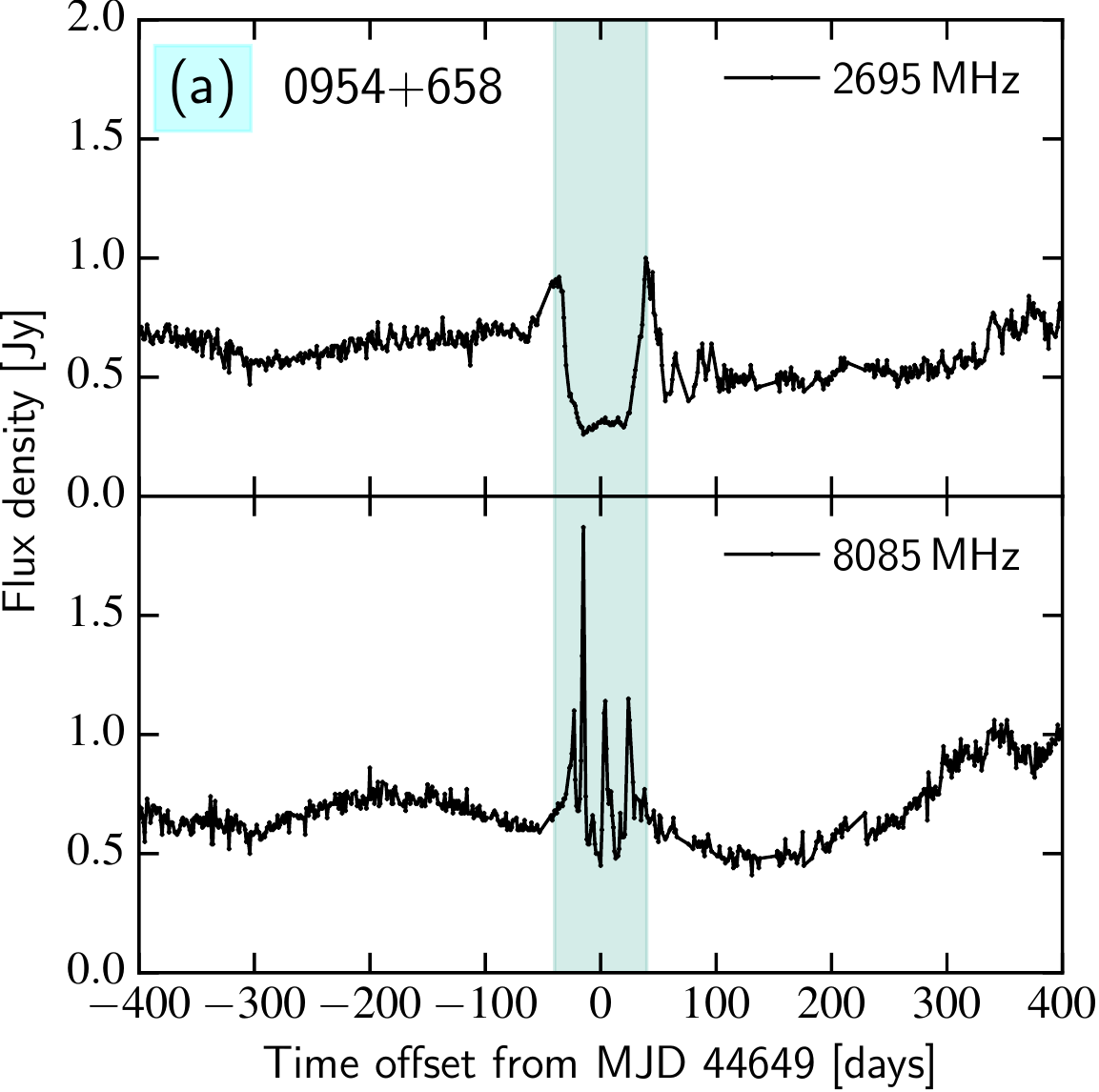} & 
\includegraphics[width=0.47\linewidth]{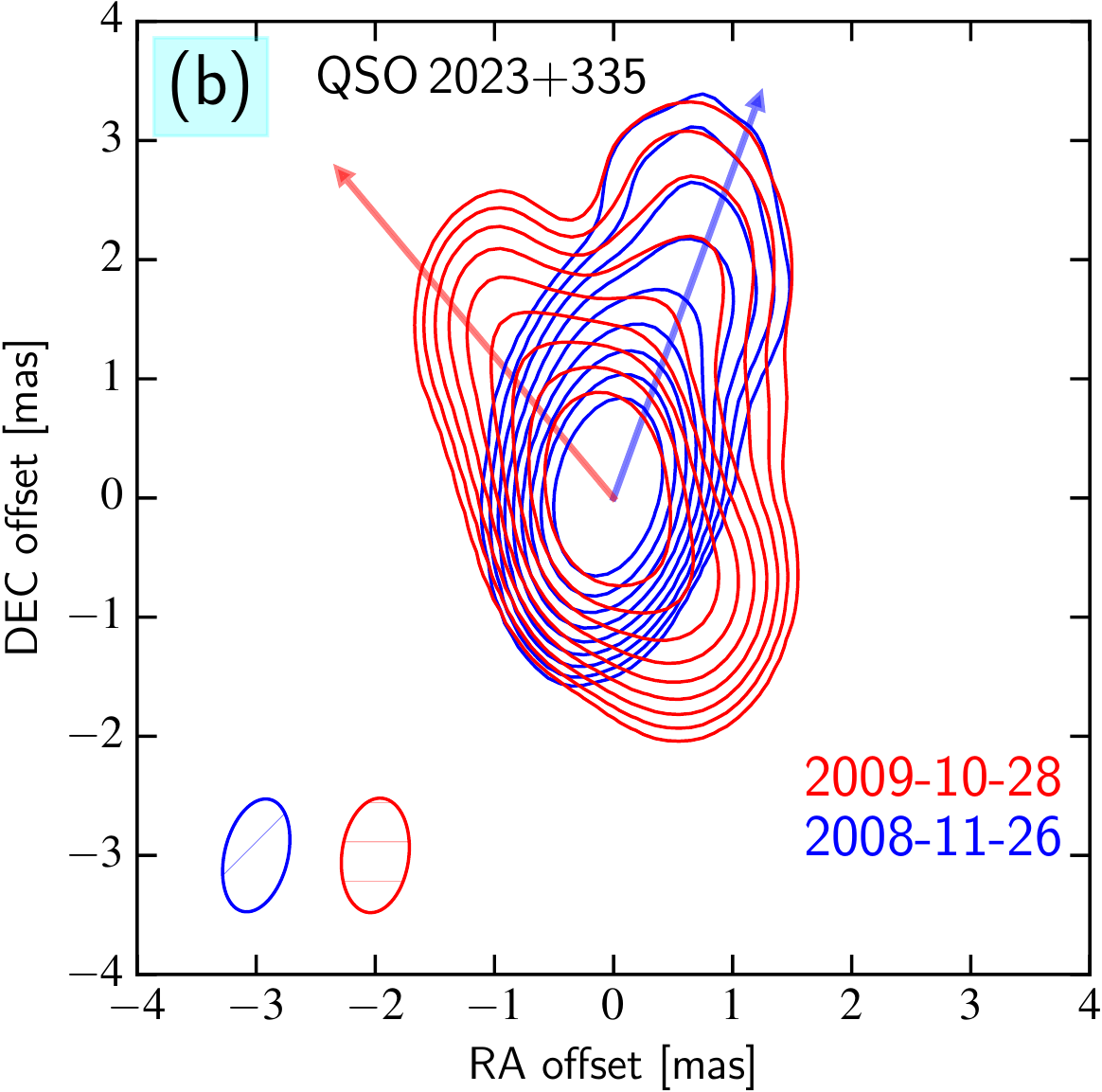}\\
\noalign{\vskip 5mm} 
\includegraphics[width=0.47\linewidth]{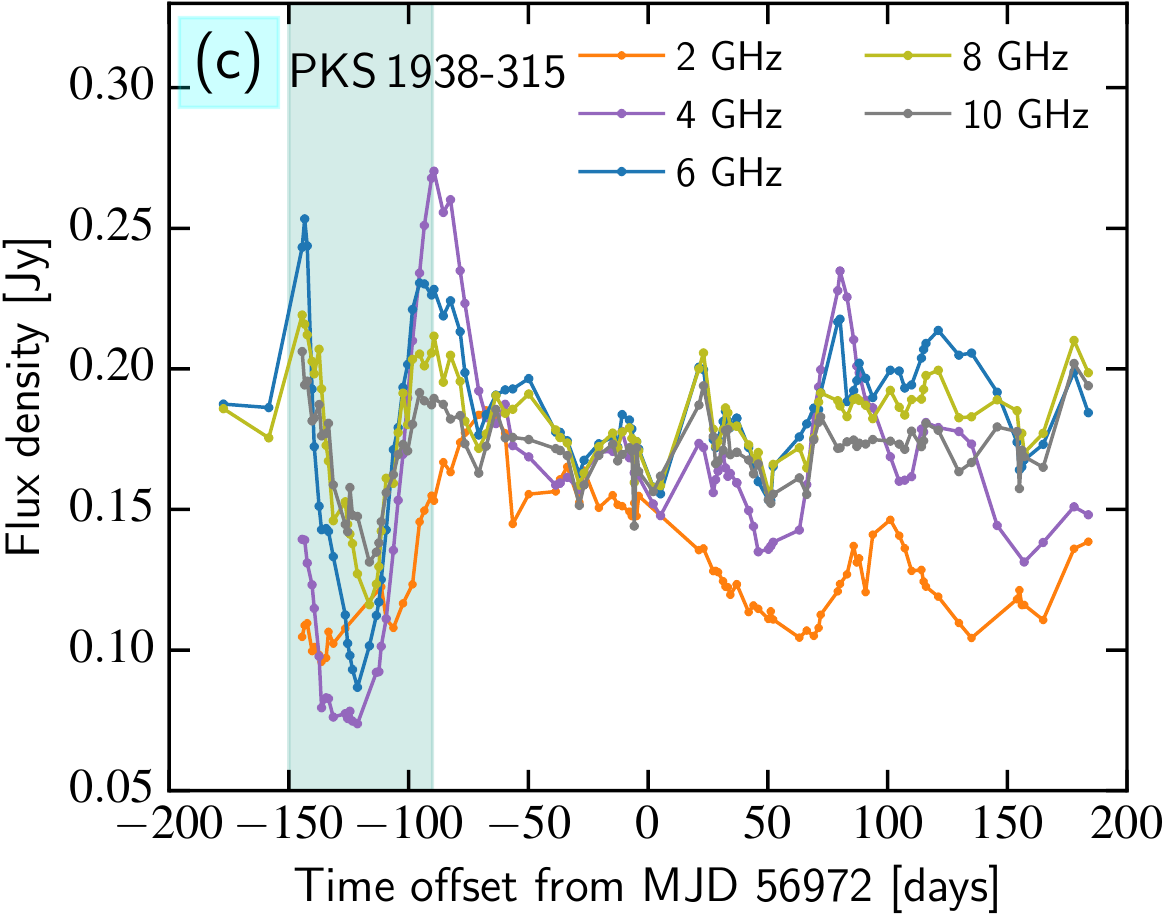} &
\includegraphics[width=0.47\linewidth]{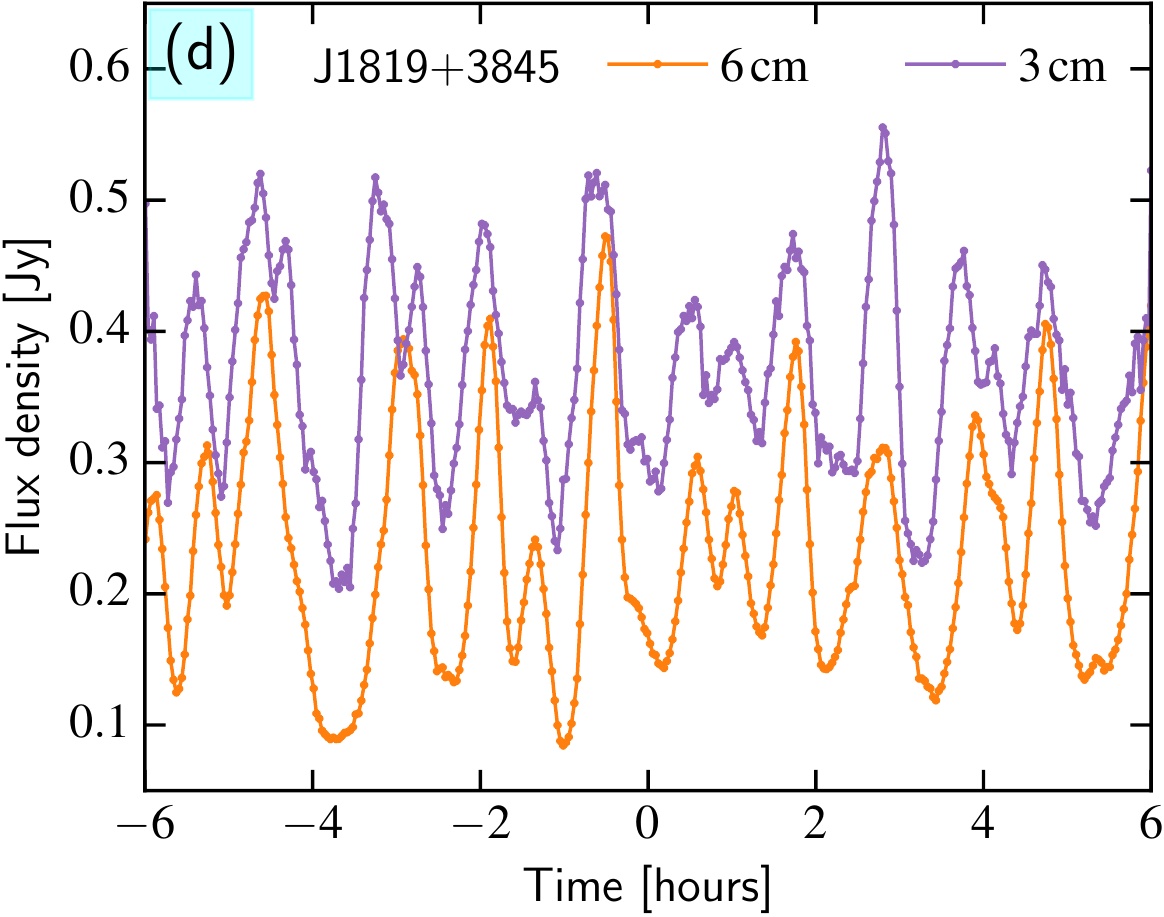}   
\end{tabular}
\caption{Observational manifestations of extreme interstellar scattering. (a) discovery of ESEs in QSO 0954+658 showing highly achromatic and time-symmetric U-shaped features \citep{fiedler1987}. (b) VLBI images of QSO 2023+335 during (red) and prior (blue) to an ESE observed in its 15\,GHz light curve (not shown for brevity), clearly showing the presence of multiple lensed images along an axis (red arrow) not aligned with the jet axis (blue arrow) \citep{pushkarev-2013,mojavecat}. (c) Real-time wide-band detection of an ESE in PKS 1938-315 that ruled out converging lens models \citep{bannister2016}. (d) Extreme scintillation in J1819+3845 ($\sim 30\,$\% flux-density modulation on timescales of tens of minutes caused by plasma clouds with density $n_e\gg 10^2\,$cm$^{-3}$ within $1-3$\,pc from the Sun \citep{j1819cat}.}
\end{figure*}
\subsection{Introduction and phenomenology}
Extreme scattering events (ESE) were first identified by \citet{fiedler1987} as rare ($\approx 7\times 10^{-3}\,$src$^{-1}\,{\rm yr}^{-1}$) $\sim\,$month-long `U'-shaped dips in radio light curves of some sources monitored by the Green bank survey (Fig 1a). The dip is bracketed by two high amplification spikes reminiscent of caustic crossing from gravitational lensing parlance. ESEs are highly chromatic due to changes in source size as well as the scattering properties of the intervening medium with spectral frequency (Fig 1a, 1c). The time-symmetry and singular nature of these events belied explanations based on intrinsic variability. Instead, transverse passage of a small, dense `plasma lens' was identified as the cause \citep{fiedler1987,romani1987}. The refractive index of plasma varies with wavelength as $\lambda^2$ explaining the frequency structure, and ray tracing through a simple Gaussian lens profile can satisfactorily explain the light curve phenomenology. The anticipated variations in dispersion measures during an ESE \citep{coles2015,kerr2018}, as well as multiple lensed images of a quasar during an ESE have been detected (Fig 1b) firmly establishing the plasma lensing hypothesis. A closely related phenomenon is extreme scintillation (or intra-day variability), the most poignant example of which was seen in J1819+3845 (Fig 1d). Though not a singular event like an ESE, the length-scale and electron densities in plasma structures that cause such scintillation is akin to that required for ESEs.\footnote{Ideally one should use the phrase ESE for extreme scintillation (J1819+3845) and call the U-shaped events as extreme lensing events (ELE), but we have retained the historic naming convention here.} We now show that the inferred ESE plasma properties are quite dramatic.

\subsection{A new component of the ISM?}
A plasma cloud of radius $r$ and electron column $N_e$ advances the phase of a electromagnetic wave of wavelength $\lambda$ by $\phi = \lambda r_e N_e$ where $r_e$ is the classical electron radius. This causes refraction of light through an angle 
$\theta_s\approx \lambda^2 N_e r_e/(2\pi r)$.
The ratio of the refraction angle to that subtended by the lens, $\alpha = D\theta_s/r$ ($D$ is the lens distance) is a measure of the lensing strength. For the cloud to have an observable impact, $\theta_s$ must be of the same order as the cloud-size ($\alpha\gtrsim 1$). To appreciate the lensing strength of ESE clouds, consider a typical ESE that last about $\sim 1$\,month which for typical Galactic proper motion of $v\sim 100$\,km\,s$^{-1}$ sets the characteristic cloud size of order $r\sim 1\,$AU. The electron density in the lens must therefore be
\begin{equation}
n_e=\frac{N_e}{r} \sim 10^3\,{\rm cm}^{-3}\,\left(\frac{r}{{\rm AU}}\right)\,\left(\frac{\lambda}{10\,{\rm cm}}\right)^{-2}\,\left(\frac{D}{1\,{\rm kpc}}\right)^{-1},
\end{equation}
with a column density of 
\begin{equation}
N_e\sim 10^{16}\,{\rm cm}^{-2}\left(\frac{r}{{\rm AU}}\right)^2\,\,\left(\frac{\lambda}{10\,{\rm cm}}\right)^{-2}\,\left(\frac{D}{1\,{\rm kpc}}\right)^{-1}.
\end{equation}
The gas pressure in the lens is over four orders of magnitude larger than the ambient ISM pressure ($\sim 0.1\,$cm$^{-3}$ in the warm ionized medium). This pressure-support problem can be circumvented in two ways: (a) an edge on sheet-like geometry can provide the required column, but extreme sheet axial ratios of order $10^4$ are required, or (b) the lenses could be ionized sheaths of self-gravitating $\sim 10^2\,$AU-scale clouds of (presumably) molecular gas (similar to cometary knots in the Helix nebula \citep{helix}; see also \citep{walker2017}). Reconciling the observed ESE-rate requires these knots to have a total mass that rivals the Galactic stellar mass--- an exotic, if not unpalatable conclusion! 
\subsection{ESEs with the ngVLA}
The molecular knots hypothesis can be tested via absorption spectroscopic signatures from associated dust and/or molecular ro-vibrational lines during an ESE. On the other hand, the non-axisymmetric lensing potential of edge-on sheets can potentially be discriminated from axi-symmetric models by multi-epoch position and morphology of lensed images with VLBI. Implementing these in practice requires a sample of ESEs with real-time detection for multi-wavelength and $\lesssim 1$\,mas resolution VLBI follow-up (see e.g. \citep{bannister2016}.  VLBI detection is crucial since the lensing strength and source size (relative to the lens) are highly degenerate.\footnote{Current non-parametric lens modeling techniques typically assume a point-like source and absence of caustic formation (see e.g. \citet{tuntsov2016}).} A weaker lens occulting a smaller source can produce similar light curves to a stronger lens (even by factor of $\sim 10^2$) occulting a relatively larger source \citep{vedantham-ese}. The main discriminant then becomes the ray deflection angles that scale linearly with lensing strength and can be readily distinguished by VLBI observations of multiple images expected on $\sim 1\,$mas scales.

Adopting the ESE rate of $7\times 10^{-3}\,$src$^{-1}\,{\rm yr}^{-1}$ \citep{fiedler1987} and observational parameters from the successful ESE survey of \citet{bannister2016}, we foresee monitoring observations of $\sim 3000$ compact flat spectrum radio quasars from the CRATES catalog \citep{crates} with measured spectral index $>-0.4$, declination north of $-30^\circ$ and 3.6\,cm flux-density in excess of 100\,mJy. With 70 subarrays of three elements each, the flux-density of such sources can be measured in each 64\,MHz channels (for ascertaining spectral modulations indicative of ESEs) between $\approx 4$ and $20$\,GHz (Band 2 and 3), to better than 30$\sigma$ in just 30\,s integrations. Assuming an overhead of 30\%, all sources can be observed within two hours. Hence an ngVLA ESE survey will yield a $\gtrsim 1$ high quality ESEs for VLBI and multi-wavelength follow-up each month with just 0.6\% of time commitment. As the target sources are bright ($\gtrsim 0.1\,$Jy), follow-up observations are feasible with existing VLBI infrastructure (VLBA and/or EVN).

\section{Symmetric achromatic variations}
\begin{figure*}
\centering
\includegraphics[width=0.8\linewidth]{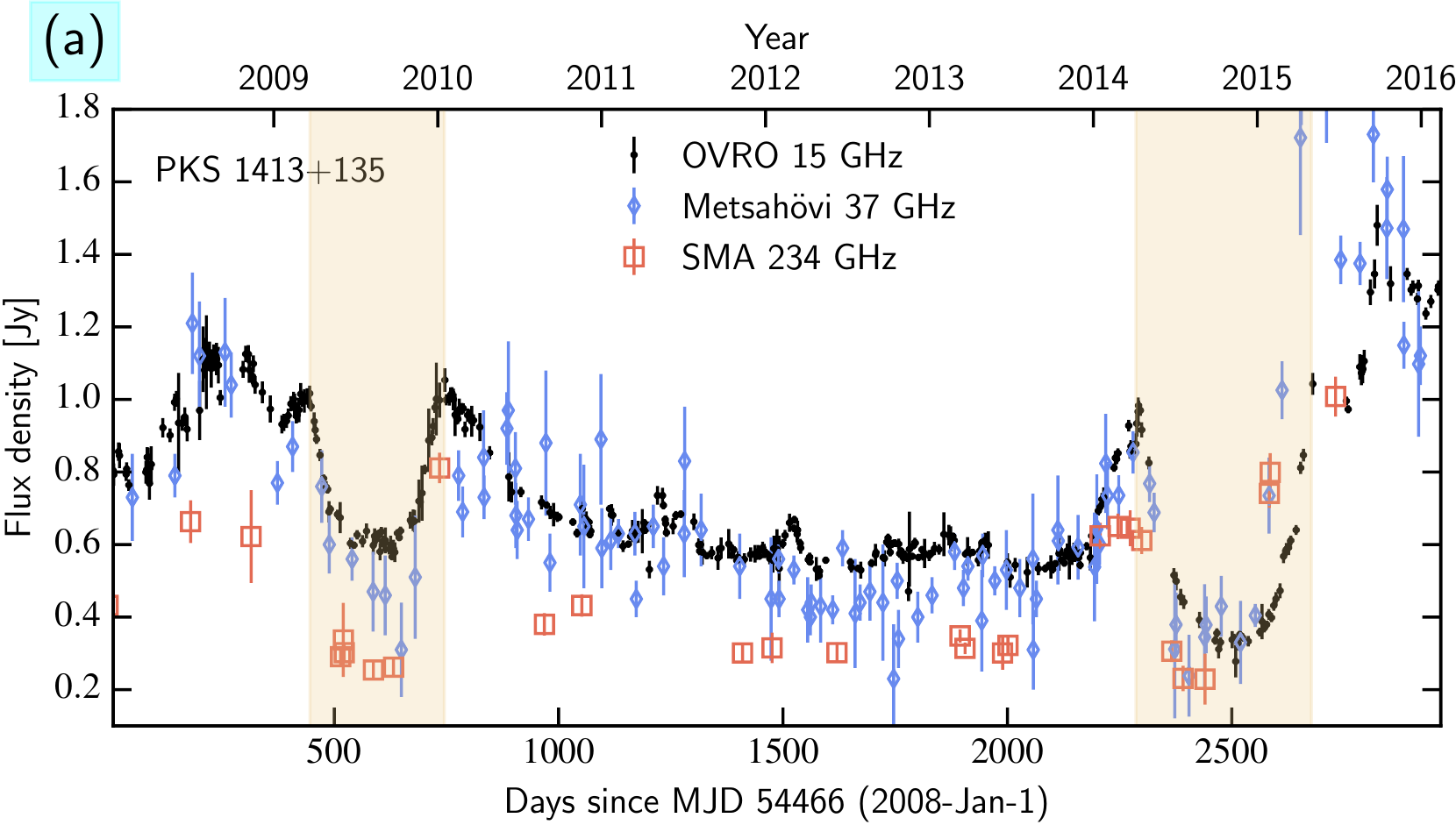}\\
\includegraphics[width=\linewidth]{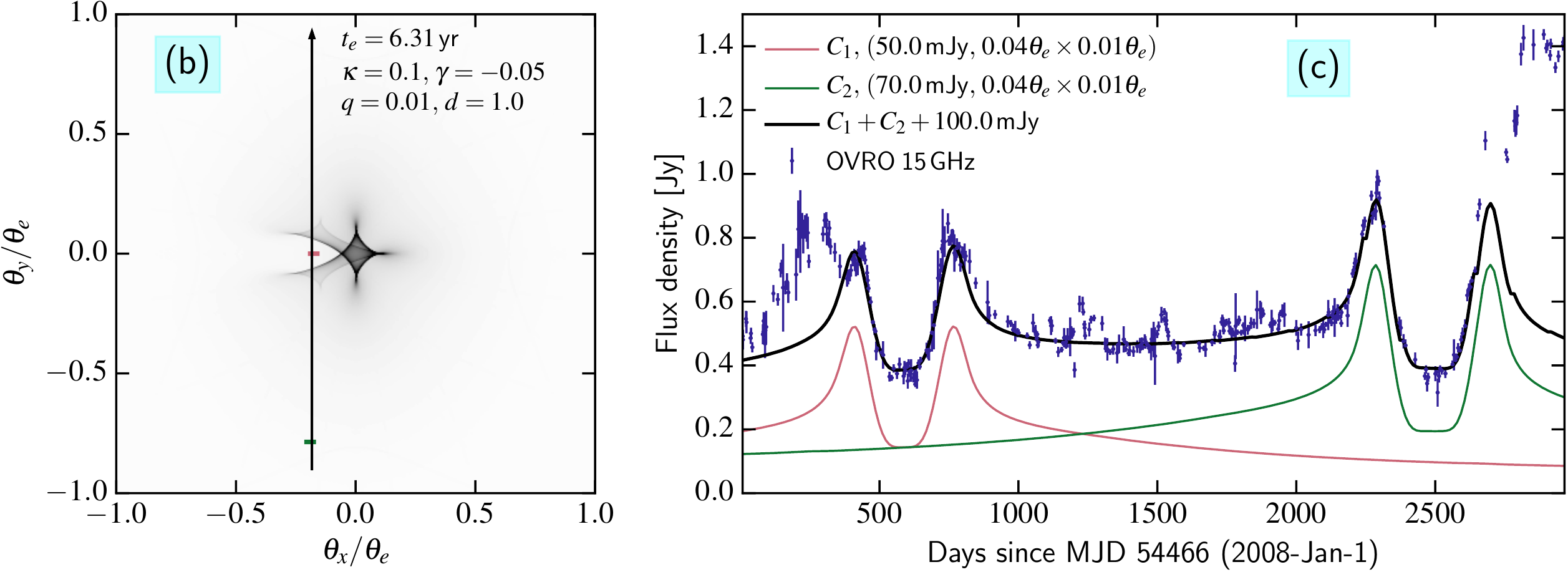}
\caption{Symmetric achromatic variations in PKS1413+135 \citep[][\copyright AAS. reproduced with permission]{vedantham-sav}. (a) light curve showing time symmetric U-shaped features that are achromatic from 15\,GHz to 234\,GHz. (b) Magnification map of a possible lens model with two point like masses ($q,\,d$: mass ratio and separation) with and external smooth mass sheet from the host galaxy mass ($\kappa,\,\gamma$: convergence and shear parameter), and path of two relativistically moving source components that can reproduce the data. (c) light curves of the two components (red and blue) and the total light curve fit to the 15\,GHz data (blue points; linear trend removed). The data also admits oblique source trajectories and other lens configurations (see Fig. 4 of \citet{vedantham-sav}).}
\end{figure*}
\subsection{Introduction and phenomenology}
A recent addition to the mystery of U-shaped light curves is the phenomenon of symmetric achromatic variations (SAV) reported by \citet{vedantham-sav}. SAV was identified in the OVRO 40\,m blazar monitoring survey, as a time-symmetric year-long U-shaped feature in the light curve of PKS1413+135 (Fig 2a). The light-curve profile is reminiscent of ESEs, but the phenomenon is observed to be achromatic from 15\,GHz to 234\,GHz--- a spectral span that constitutes a factor of $\sim 400$ in plasma lensing strength. This makes the ESE hypothesis highly problematic \citep{vedantham-ese} and \citet{vedantham-sav} considered gravitational lensing (which is achromatic) as a cause. The light curve can be fit with a physical lens model that consists of two point-like masses with mass ratio $\sim 0.01$ in the presence of an external mass sheet presumably from the galaxy that hosts the point-like masses (Fig 2b, 2c). Radio sources such as PKS1413+135 are known to have relativistically moving compact ($\sim \mu$as-scale) components. The lensing magnification modulates the flux-density of these components as they cross the lens yielding the observed light curve. The phenomena is similar to optical microlensing of Galactic stars \citep{udalski1993,alcock1993} but probes a significantly larger mass-scale at cosmic distances. For instance, in case of PKS1413+135, the lens mass is constrained to lie between $10^3$ and $10^6\,M_\odot$ depending on the precise lensing geometry. Finally, for strong lensing to be viable, the lenses must be particularly compact ($\gtrsim 10^3\,M_\odot\,{\rm pc}^{-2}$). Examples of objects with such compactness include globular clusters, dense molecular cores, and certain dark matter candidates including primordial black holes (see e.g. \citet{carr2017}).

\subsection{A population of intermediate mass lenses?}
Gravitational lensing leads to multiple imaging on angular scales of $\Delta\theta\approx\theta_e$ where the Einstein radius, $\theta_e$ depends on the lens mass $M_l$ and the lensing geometry: \[\theta_e \approx 90\,\upmu{\rm as}\left(\frac{M_l}{10^3\,M_\odot} \right)^{1/2}\left(\frac{D_lD_s/D_{ls}}{1{\rm Gpc}}\right)^{-1/2}.\] where $D_l$, $D_s$ and $D_{ls}$ are the observer$-$lens, observer$-$source and lens$-$source distance respectively. To obtain a distinct SAV signature, the source must be no larger than a few percent of the Einstein radius. The angular size of compact steep-spectrum sources is constrained by inverse Compton losses: their brightness temperatures are bound to be  $\lesssim 10^{12}\,$K, while relativistic effects can lead to apparent brightness temperature of $\sim 10^{13}\,$K \citep{johnson2016}. Observations of SAV therefore places a lower bound on the Einstein radius and hence places the mass-scale of the putative lens in the intermediate mass regime: \[M_{\rm l}\gtrsim 10^3\,M_\odot \left( \frac{S}{0.1\,{\rm Jy}}\right)\  \left( \frac{\lambda}{1\,{\rm cm}}\right)^2\left(\frac{D_lD_s/D_{ls}}{1{\rm Gpc}}\right)\]

The compact $\mu$as-scale components (unresolved by VLBI) typically have relativistic velocities and will cross such a lens in a timescale of \[t \approx 1.4\,{\rm yr} \left(\frac{M_l}{10^3\,M_\odot} \right)^{1/2}\left(\frac{D_lD_s/D_{ls}}{1{\rm Gpc}}\right)^{1/2}.\]
This underscores the specific utility of relativistic radio sources. While typical virial speeds of $\sim 300$\,km\,s$^{-1}$ can only cross a $10^3\,M_\odot$ lens in time-scales of centuries, relativistic sources can cross on $\sim$\,year-long timescales which allows for detection of lensing candidates with flux-density monitoring campaigns alone. We note that, while we are not aware of any other mechanism that can explain SAVs, the lensing hypothesis must await confirmation from detection of lensed images in VLBI follow-up of future SAVs. Though challenging, observations with existing global mm-wave VLBI techniques have the required resolution of several tens of $\mu$as. 

\subsection{SAVs and the ngVLA}
The main focus to solve the SAV puzzle is in real-time identification of an SAV for mm-wave VLBI follow-up. SAVs are rare: the events in PKS1413+135 were found from a sample of $10^3$\,sources $\times 8$\,yr of data (cadence of twice per week). Moreover, as in the case of PKD1413+135, light curves at widely spaced frequencies are required to discriminate SAVs from ESEs. bearing these in mind, we foresee an ngVLA monitoring survey of about $8000$ flat-spectrum sources ($\alpha>-0.5$) north of $-30^\circ$\,declination in the the CRATES catalog once every fortnight\footnote{With a lack of chromatic signatures, time-symmetry is key to differentiating intrinsic variations and, if possible a cadence of one per week is ideal.}, in bands 3 and 5, to find $\sim 1$ year-long SAV candidate per year for multi-epoch mm-wave VLBI follow-up. We end by noting that confirmation of gravitational lensing as a cause of SAVs would open up a new sub-field of study into cosmic matter on small mass-scales which can potentially have important implication for evidencing the nature of dark matter.

\bibliography{refs}
\end{document}